\documentclass[jap,reprint,showpacs]{revtex4-1}
\usepackage{dcolumn}
\usepackage{bm}
\usepackage{amsmath}
\usepackage{amssymb}
\usepackage{revsymb}
\usepackage{graphics}
\usepackage{graphicx}
\usepackage{natbib}
\usepackage{color}

\begin{document}

\title{Radiative energy and momentum transfer for various spherical shapes: a single sphere, a bubble, a spherical shell and a coated sphere}
\author{Yi Zheng}
\email{zheng@egr.uri.edu}
\affiliation{ 
Department of Mechanical, Industrial and Systems Engineering, University of Rhode Island, Kingston, RI 02879, USA
}
\author{Alok Ghanekar}
\affiliation{ 
Department of Mechanical, Industrial and Systems Engineering, University of Rhode Island, Kingston, RI 02879, USA
}

\begin{abstract}
We use fluctuational electrodynamics to determine emissivity and van der Waals contribution to surface energy for various spherical shapes, such as a sphere, a bubble, a spherical shell and a coated sphere, in a homogeneous and isotropic medium. Near-field radiative transfer and momentum transfer between flat plates and curved surfaces have been studied for the past decades, however the investigation of radiative heat transfer and van der Waals stress due to fluctuations of electromagnetic fields for a single object is missing from literature. The dyadic Green's function formalism of radiative energy and fluctuation-induced van der Waals stress for different spherical configurations have been developed. We show (1) emission spectra of micro and nano-sized spheres display several emissivity sharp peaks as the size of object reduces, and (2) surface energy becomes size dependent due to van der Waals phenomena when size of object is reduced to a nanoscopic length scale.

\end{abstract}

\maketitle

\section{Introduction}
Quantum and thermal fluctuations of electromagnetic fields, which give rise to Planck's
law of blackbody radiation \cite{planck2011theory}, are responsible for radiative energy, momentum and entropy transfer \cite{zheng2014near,basu2009review,zheng2014patch,francoeur2011electric,narayanaswamy2013theory}. Classical theory of thermal radiation takes only propagating waves into consideration \cite{planck2011theory}. Presence of evanescent waves nearby the surfaces leads to near-field effects, such as interference, diffraction and tunneling of surface waves, which are important when the size and length scale is comparable to the thermal wavelength of objects ($\approx$ 10 $\mu$m at room temperature 300 K) \cite{zheng2014near}. Enhancement of radiative energy transfer, wavelength selectivity, and size and scale dependence are typical characteristics of near-field phenomena which make it more interesting. These characteristics could be exploited in many potential nanotechnological applications such as nanoparticles \cite{perez2008heat,huth2010shape,von1999preparation,biehs2010near} and wavelength selective absorber and emitters \cite{rephaeli2009absorber,avitzour2009wide,sergeant2010high,cao2010semiconductor,narayanaswamy2014infrared}. Fluctuations in electromagnetic fields due to the presence of boundaries create van der Waals pressures which lead to momentum transfer \cite{belosludov1975contribution,pitaevskii2006comment,narayanaswamy2013van,parsegian2006van}. These phenomena can be described by cross-spectral densities of the electromagnetic field using dyadic Green's functions of vector Helmholtz equation \cite{tai1994dyadic,narayanaswamy2013green,yla2003efficient}. Such a relation between near-field radiative energy and momentum transfer for arbitrarily shaped objects has been explicitly derived in Ref. \cite{narayanaswamy2013green}.

\begin{figure*}
\centering
\includegraphics[width=14 cm]{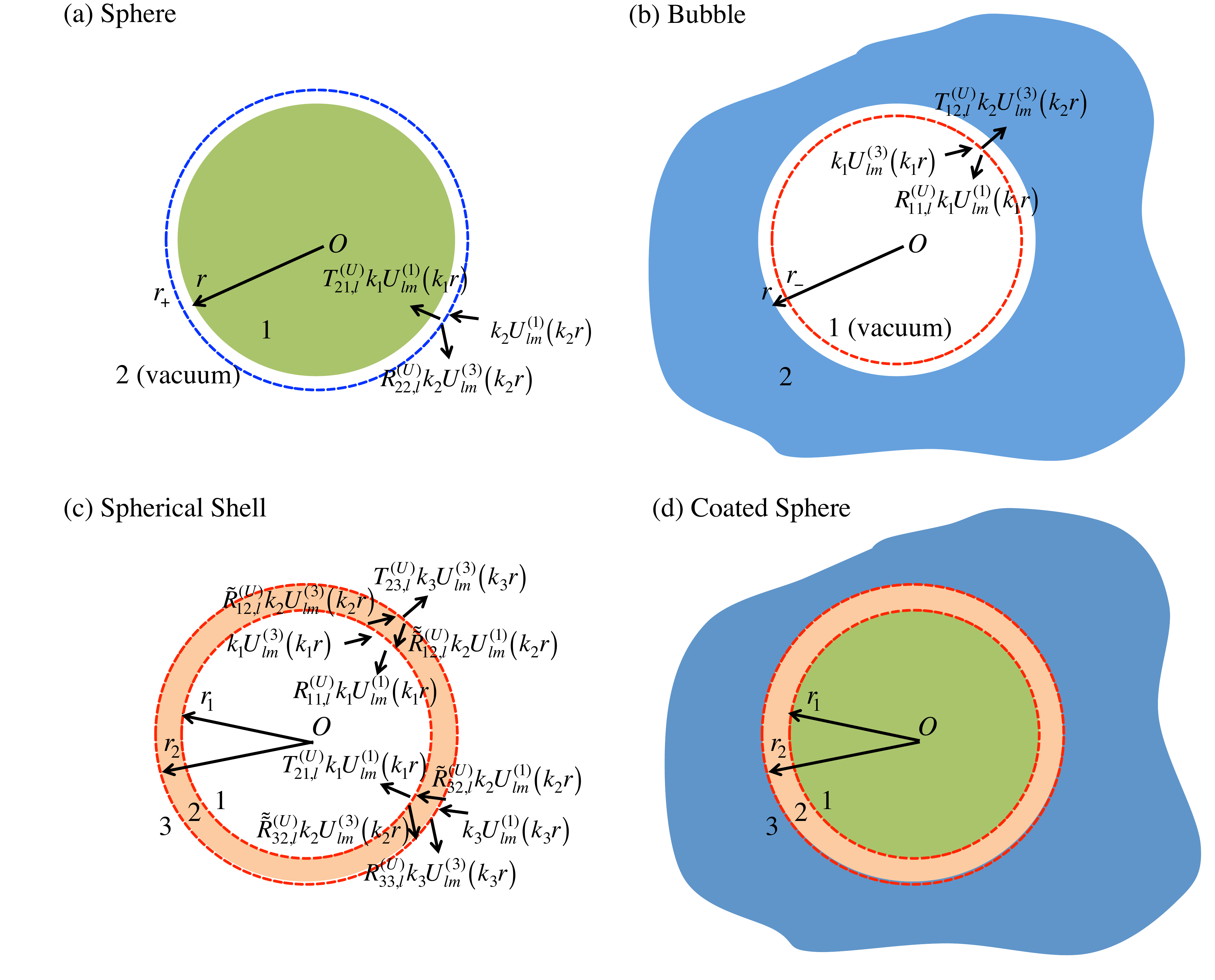}
\caption{\label{fig:schematic} Schematic of various spherical geometries. (a) a sphere of radius $r$ (material 1, green), (b) a bubble of radius $r$ in material 2 (blue), (c) a spherical shell of interior and exterior radii $r_1$ and $r_2$ (material 2, orange), and (d) a sphere of radius $r_1$ (material 1, green) coated with a thin film of thickness $r_2-r_1$ (material 2, orange) in material 3 (blue). The short arrows on either side of the interface between two materials denote the spherical electromagnetic waves $U$($\bm{M}$ and $\bm{N}$ waves). $U_{lm}^{(p)}$ is vector spherical wave functions of order $(l,m)$, and superscript $(p)$ refers to the radial behavior of the waves. $k_i$ is wavenumber in medium $i$, $R_{ij,l}^{(U)}$ and $T_{ij}^{(U)}$ are the Mie reflection and transmission coefficients due to spherical electromagnetic waves $U$ from medium $i$ to medium $j$, respectively. 
}
\end{figure*}

During the past decades, many theoretical works have been published on the topic of the enhanced thermal radiation due to near-field effects and fluctuation-induced van der Waals pressures. Most of them are focused on the configurations, for example, between planar multilayered media \cite{ottens2011near,francoeur2009solution,dzyaloshinskii1961general,narayanaswamy2013van,parsegian1973van,antezza2008casimir}, cylindrical objects \cite{sheikholeslami2014unsteady,edalatpour2014thermal}, spheres \cite{perez2008heat,sasihithlu2011proximity,zhu2013ultrahigh,narayanaswamy2008near,incardone2014heat}, and a sphere and a flat plate \cite{otey2011numerically,biehs2010near,otey2014fluctuational,incardone2014heat}. Although considerable amount of work devoted to theoretical and experimental results for near-field radiative energy transfer and momentum transfer between objects of the abovementioned typical geometries, these phenomena for a single object have not been studied well to the best of our knowledge. In this paper, emissivity of thermal radiation and van der Waals contribution to surface energy are determined for micro and nano-sized spherical objects of interest, such as a single sphere, a bubble, a sherical shell and a coated sphere, as shown in Fig. \ref{fig:schematic}. 

This paper demonstrates how fluctuational electrodynamics can be used to determine emissivity and van der Waals contribution to surface energy for various shapes in a homogenous and isotropic medium. The dyadic Green's function formalism of radiative heat transfer and van der Waals stress for different spherical configurations has been developed. Size dependence and wavelength selectivity of micro and nano-sized spheres are clearly observed. For small sized objects, the emissivity spectra show sharp peaks at wavelengths corresponding to characteristics of the material's refractive index. The peaks increase as size of object decreases. This is consistent with emissivity calculations done for thin film as in Ref. \cite{narayanaswamy2014infrared}. The calculations for a spherical bubble show that as radius of the bubble reduces, van der Waals energy contribution to the total surface energy increases and dominates it beyond a critical value. 

The paper is arranged as follows. Section \ref{sec:fundamental} focuses on the theoretical formulations for near-field radiative energy transfer and momentum transfer for a generalized case. Calculations of transmissivity for radiative energy transfer and fluctuation-induced van der Waals pressure are discussed in Sec. \ref{subsec:transEnergy} and Sec. \ref{subsec:transMomentum}, respectively. Using this formalism, the analytical expressions of Mie reflection and transmission coefficients for specific cases of a sphere, a bubble, a spherical shell and a coated sphere in a homogeneous and isotropic medium are obtained in Sec. \ref{sec:discussion}. Formulae for calculations of emissivity and van der Waals contribution to surface energy of a spherical body are subsequently derived in Sec. \ref{subsec:emissivity} and Sec. \ref{subsec:vdW}.  We recapitulate the paper and states the conclusions in Sec. \ref{sec:sum}.

\section{\label{sec:fundamental}Theoretical fundamentals}

The Fourier transforms of the electric and magnetic fields due to thermal sources in any object (volume $V_s$) are given by \cite{rytov1967theory,narayanaswamy2013green}
\begin{subequations}
\label{eqn:fieldrelations}
\begin{equation}
\label{eqn:efieldeqn}
\mathbf{E}(\mathbf{\tilde{r}}) =\int\limits_{V_s} [ \mathbf{p} (\mathbf{r}) \cdot \overline{\overline{\mathbf{G}}}_{e}(\mathbf{r},\mathbf{\tilde{r}}) - \mathbf{J}^{m}(\mathbf{r}) \cdot \overline{\overline{\mathbf{G}}}_{E}(\mathbf{r}, \mathbf{\tilde{r}}) ]d\mathbf{r}
\end{equation} 
\begin{equation}
\label{eqn:mfieldeqn}
\mathbf{H}(\mathbf{\tilde{r}})= \int\limits_{V_s} [\mathbf{m} (\mathbf{r}) \cdot \overline{\overline{\mathbf{G}}}_{m}(\mathbf{r}, \mathbf{\tilde{r}}) + \mathbf{J}^{e}(\mathbf{r}) \cdot \overline{\overline{\mathbf{G}}}_{M}(\mathbf{r}, \mathbf{\tilde{r}}) ] d\mathbf{r}
\end{equation}
\end{subequations}

\noindent where source position vector $\mathbf{r}\in V_s$, observation position vector $\mathbf{\tilde{r}}$, $\mathbf{J}^{e}$ and $\mathbf{J}^{m}$ are the electric and magnetic current densities, $\mathbf{p} \left(\mathbf{r}\right)$$=$$i\omega\mu_{o}\mu(\mathbf{r})\mathbf{J^{e}}\left(\mathbf{r}\right)$, $\mathbf{m} \left(\mathbf{r}\right) = i\omega\varepsilon_{o}\varepsilon(\mathbf{r}) \mathbf{J^{m}}\left(\mathbf{r}\right)$, 
$ \varepsilon_o $ and $ \mu_o $ are the permittivity and permeability of free space, and, $ \varepsilon(\mathbf{r}) $ and $ \mu(\mathbf{r}) $ are the permittivity and permeability at location $\mathbf{r}$. $\overline{\overline{\mathbf{G}}}_{E}(\mathbf{r},\mathbf{\tilde{r}})=\nabla \times \overline{\overline{\mathbf{G}}}_{e}(\mathbf{r},\mathbf{\tilde{r}})$ and $\overline{\overline{\mathbf{G}}}_{M}(\mathbf{r},\mathbf{\tilde{r}}) =\nabla \times \overline{\overline{\mathbf{G}}}_{m}(\mathbf{r},\mathbf{\tilde{r}})$. $\overline{\overline{\mathbf{G}}}_{e}(\mathbf{r},\mathbf{\tilde{r}}) $ and $\overline{\overline{\mathbf{G}}}_{m}(\mathbf{r},\mathbf{\tilde{r}}) $ are dyadic Green's functions of the vector Helmholtz equation that satisfy the following boundary conditions on the interface between object 1 and object 2. For electric dyadic Green's functions,
\begin{subequations}
\begin{equation}
\label{eqn:ebc}
\hat{\mathbf{n}}_1 \times \mu_1 \overline{\overline{\mathbf{G}}}_{e}(\mathbf{r}_1,\mathbf{\tilde{r}}) = \hat{\mathbf{n}}_2 \times \mu_{2} \overline{\overline{\mathbf{G}}}_{e}(\mathbf{r}_{2},\mathbf{\tilde{r}}), 
\end{equation}
\begin{equation}
\label{eqn:Ebc}
\hat{\mathbf{n}}_1 \times \overline{\overline{\mathbf{G}}}_{E}(\mathbf{r}_1,\mathbf{\tilde{r}}) = \hat{\mathbf{n}}_2 \times \overline{\overline{\mathbf{G}}}_{E}(\mathbf{r}_{2},\mathbf{\tilde{r}}),
\end{equation}
and for magnetic dyadic Green's functions,
\begin{equation}
\label{eqn:mbc}
\hat{\mathbf{n}}_1 \times \varepsilon_1 \overline{\overline{\mathbf{G}}}_{m}(\mathbf{r}_1,\mathbf{\tilde{r}}) = \hat{\mathbf{n}}_2 \times \varepsilon_{2} \overline{\overline{\mathbf{G}}}_{m}(\mathbf{r}_{2},\mathbf{\tilde{r}}), 
\end{equation}
\begin{equation}
\label{eqn:Mbc}
\hat{\mathbf{n}}_1 \times \overline{\overline{\mathbf{G}}}_{M}(\mathbf{r}_1,\mathbf{\tilde{r}}) =\hat{\mathbf{n}}_2 \times \overline{\overline{\mathbf{G}}}_{M}(\mathbf{r}_{2},\mathbf{\tilde{r}}),
\end{equation}
\end{subequations}
where $\hat{\mathbf{n}}_1$ and $\hat{\mathbf{n}}_2$ are surface normal vectors, and $\mathbf{r}_1$ and $\mathbf{r}_{2}$ are position vectors of points on either side of interface in volume $V_1$ and $V_2$, as $\lvert \mathbf{r}_1- \mathbf{r}_2 \rvert \rightarrow 0 $. 

\subsection{\label{subsec:transEnergy}Transmissivity for radiative energy transfer}

The steady state radiative heat transfer from object $ 1 $ to object $ 2 $, $ \mathcal{Q}_{1 \rightarrow 2} $, is given by
\begin{equation}
\label{eqn:Q12surfintegral}
\mathcal{Q}_{1 \rightarrow 2} = -\oint\limits_{S_2} \mathbf{P}(\mathbf{\tilde{r}}) \cdot \hat{\mathbf{n}}_2 d\mathbf{\tilde{r}} 
\end{equation}
where $\mathbf{P}(\mathbf{\tilde{r}})=\langle \mathbf{E}(\mathbf{\tilde{r}},t) \times \mathbf{H}(\mathbf{\tilde{r}},t) \rangle$ is the Poynting vector at $\mathbf{\tilde{r}} \in S_2$ due to thermally fluctuating sources within $V_1$. $\langle$ $\rangle$ denotes the ensemble average, and the ``$ - $'' sign in front of the surface integral is because $ \hat{\mathbf{n}}_2 $ is the outward pointing normal on the surface $ S_2 $. The components of $\mathbf{P}(\mathbf{\tilde{r}})$ are given by
\begin{equation}
\label{eqn:poyntingcomponent}
P_i(\mathbf{\tilde{r}})=\epsilon_{ipq} \langle E_p(\mathbf{\tilde{r}},t) H_q(\mathbf{\tilde{r}},t) \rangle
\end{equation}
where $p$, $q$=1,2,3 are the labels for the Cartesian components of the vector, $\epsilon_{ipq}$ is the Levi-Civita symbol, $ \langle E_p(\mathbf{\tilde{r}},t) H_q(\mathbf{\tilde{r}},t) \rangle $ is the contribution to $ \langle E_p(\mathbf{\tilde{r}},t) H_q(\mathbf{\tilde{r}},t) \rangle $ from sources within $ V_1 $. We note that $\displaystyle \langle E_q(\mathbf{\tilde{r}},t) H_p(\mathbf{\tilde{r}},t) \rangle = \int\limits_0^{\infty} \frac{d\omega}{2\pi} \langle E_q(\mathbf{\tilde{r}},\omega) H_p^*(\mathbf{\tilde{r}},\omega) \rangle$. Using Eq. \ref{eqn:efieldeqn} and Eq. \ref{eqn:mfieldeqn}, Eq. \ref{eqn:Q12surfintegral} for $ \mathcal{Q}_{1 \rightarrow 2} $ can be re-written as
\begin{equation}
\label{eqn:T12definition}
\begin{split}
\mathcal{Q}_{1 \rightarrow 2} = \int\limits_{0}^{\infty} d\omega \frac{\hbar \omega}{2\pi} \coth\left(\frac{\hbar\omega}{2k_B T_1}\right) \mathcal{T}_{rr}(\omega) 
\end{split}
\end{equation}
where $k_B$ is Boltzmann constant, $2\pi\hbar$ is the Planck constant, $T_1$ is the absolute temperature of object 1, and $\mathcal{T}_{rr}(\omega)$ is a generalized transmissivity for radiative energy transport between objects 1 and 2. $\mathcal{T}_{rr}(\omega)$ can be expressed exclusively in terms of components of the dyadic Green's functions on the surfaces of objects 1 and 2, which is given by \cite{narayanaswamy2013green}
\begin{widetext}
\begin{equation}
\label{eqn:generalizedtransmissivity1}
\begin{split}
\mathcal{T}_{rr}(\omega) = \Re Tr \oint\limits_{S_1} d\mathbf{r} \oint\limits_{S_2} d\mathbf{\tilde{r}} 
\bigg[ \frac{\omega^2}{c^2} \left[\mathbf{\hat{n}}_2 \times \mu_2 \overline{\overline{G}}_e(\mathbf{\tilde{r}},\mathbf{r}) \right] \cdot 
\left[\mathbf{\hat{n}}_1 \times \varepsilon_1^* \overline{\overline{G}}_m^*(\mathbf{r},\mathbf{\tilde{r}}) \right] + \left[\mathbf{\hat{n}}_2 \times \overline{\overline{G}}_E(\mathbf{\tilde{r}},\mathbf{r}) \right] \cdot 
\left[\mathbf{\hat{n}}_1 \times \overline{\overline{G}}^*_E(\mathbf{r},\mathbf{\tilde{r}}) \right] \bigg] 
\end{split}
\end{equation}
\end{widetext}
where $\Re$ stands for the real part, superscript $^*$ denotes the complex conjugate, and operator trace $Tr(\overline{\overline{A}}) = \sum\limits_{p=1}^{3} A_{pp}$.

\subsection{\label{subsec:transMomentum} van der Waals stress for radiative momentum transfer}

Radiative momentum transfer arose from the fluctuations of electromagnetic fields is responsible for the van der Waals pressure, which can be determined from the electromagnetic stress tensor, $ \overline{\overline{\sigma}}=\overline{\overline{\sigma}}_e+\overline{\overline{\sigma}}_m $, where $ \overline{\overline{\sigma}}_e $ and $ \overline{\overline{\sigma}}_m $ are the electric and magnetic field contributions respectively. Stress tensors $ \overline{\overline{\sigma}}_e $ and $ \overline{\overline{\sigma}}_m $ are given by \cite{zheng2011lifshitz,narayanaswamy2013van}
\begin{subequations}
\begin{equation}
\label{eqn:efieldstresstensor}
\overline{\overline{\sigma}}_e(\mathbf{\tilde{r}}) = \varepsilon_o \left[ \langle \mathbf{E}(\mathbf{\tilde{r}},t) \mathbf{E}(\mathbf{\tilde{r}},t) \rangle -\frac{1}{2} \overline{\overline{I}} \langle \mathbf{E}^2(\mathbf{\tilde{r}},t) \rangle \right]
\end{equation}
\begin{equation}
\label{eqn:mfieldstresstensor}
\overline{\overline{\sigma}}_m(\mathbf{\tilde{r}}) = \mu_o \left[ \langle \mathbf{H}(\mathbf{\tilde{r}},t) \mathbf{H}(\mathbf{\tilde{r}},t) \rangle -\frac{1}{2} \overline{\overline{I}} \langle \mathbf{H}^2(\mathbf{\tilde{r}},t) \rangle \right],
\end{equation}
\end{subequations}
where $\overline{\overline{I}} $ is the identity matrix, $ \langle \mathbf{E}(\mathbf{\tilde{r}},t) \mathbf{E}(\mathbf{\tilde{r}},t) \rangle $ and $ \langle \mathbf{H}(\mathbf{\tilde{r}},t) \mathbf{H}(\mathbf{\tilde{r}},t) \rangle $ are matrices whose components are $ \langle E_p(\mathbf{\tilde{r}},t) E_q(\mathbf{\tilde{r}},t) \rangle $ and $ \langle H_p(\mathbf{\tilde{r}},t) H_q(\mathbf{\tilde{r}},t) \rangle $ respectively.

Relying on Rytov's theory of fluctuational electrodynamics \cite{rytov1967theory}, the cross-spectral correlations of the electric and magnetic field components can be written as \cite{narayanaswamy2013green}
\begin{subequations}
\begin{equation}
\label{eqn:Efields}
\begin{split}
\left<E_i(\mathbf{\tilde{r}},\omega)E_j^*(\mathbf{\tilde{r}},\omega)\right>= & \frac{\hbar\omega^2}{\pi}\mu_0 \coth\left(\frac{\hbar \omega}{2 k_B T}\right) \\ & \times \Im \left[\varepsilon(\omega)\mu(\omega) G_{e,ij}(\mathbf{\tilde{r}},\mathbf{\tilde{r}})\right]
\end{split}
\end{equation}
\begin{equation}
\label{eqn:Mfields}
\begin{split}
\left<H_i(\mathbf{\tilde{r}},\omega)H_j^*(\mathbf{\tilde{r}},\omega)\right> = & \frac{\hbar\omega^2}{\pi}\varepsilon_0\coth\left(\frac{\hbar \omega}{2 k_B T}\right) \\ & \times \Im\left[\varepsilon(\omega)\mu(\omega) G_{m,ij}(\mathbf{\tilde{r}},\mathbf{\tilde{r}})\right] 
\end{split}
\end{equation}
\end{subequations}
where $\Im$ denotes the imaginary part. $\varepsilon(\omega)$ and $\mu(\omega)$ are frequency dependent permittivity and permeability of region in which $\mathbf{\tilde{r}}$ is located. $\sqrt{\varepsilon(\omega)}=n(\omega)+i\kappa(\omega)$ and $\mu(\omega)=1$. The optical data for $n$ and $\kappa$ can be obtained from Refs. \cite{hough1980calculation, palik1998handbook}. The van der Waals pressure on the interior surface of a bubble or the exterior surface of a sphere of radius $r$ in an otherwise homogeneous medium (see Figs. \ref{fig:schematic}a and \ref{fig:schematic}b) can be obtained from the $\hat{r}\hat{r}$ component of the electromagnetic stress tensor, in terms of dyadic Green's functions, which is given by
\begin{widetext}
\begin{equation}
\begin{split}
\label{eqn:stensor}
\mathcal{S}_{rr}(r)&=\int\limits_0^{\infty} d\omega \frac{\hbar\omega^2}{\pi c^2}\coth\!\left(\! \frac{\hbar \omega}{2k_B T} \! \right) \Im \Bigg\{ \varepsilon(\omega)\mu(\omega) \left[G_{e,\hat{r}\hat{r}}(\bm{r},\bm{r})-\frac{1}{2} Tr \overline{\overline{G}}_{e}(\bm{r},\bm{r}) + G_{m,\hat{r}\hat{r}}(\bm{r},\bm{r})-\frac{1}{2} Tr \overline{\overline{G}}_{m}(\bm{r},\bm{r}) \right] \! \Bigg\}
\end{split}
\end{equation}
\end{widetext}
where $Tr$ is the trace of the tensor in the spherical coordinates, that is $Tr\left( \overline{\overline{G}}_{p} \right)= G_{p,\hat{r} \hat{r}}+G_{p,\hat{\theta} \hat{\theta}}+G_{p,\hat{\phi} \hat{\phi}} $. The integral on the right hand side of Eq. \ref{eqn:stensor} is of the form $
\int\limits_{0^+}^{\infty} d\omega \coth\left(\frac{\hbar \omega}{2k_B T_l}\right) \Im f(\omega) $, in which the function $f(\omega + i\xi)$ is analytic in the upper--half of the complex plane ($\xi>0$). Since $\displaystyle G\left(\bm{r},\omega \right)$ is analytic in the upper--half of the complex plane, the integral along the real frequency axis can be transformed into summation over Matsubara frequencies, $i\xi_n = in 2\pi k_B T_l/\hbar$, as $\displaystyle -\frac{2\pi k_B T}{\hbar} \sum\limits_{n=0}^{\infty}\phantom{}^{'} f(i\xi_n)$. $\sum\limits_{n=0}^{\infty}\phantom{}^{'}$ indicates that the $n=0$ term is multiplied by 1/2 \cite{lifshitz1956theory}.

$\mathcal{S}_p\left(r \right) $ can be split into two parts: (1) a part which arises from the dyadic Green's function in an infinite homogeneous medium, and (2) a part which arises from the scattered dyadic Green's function due to the presence of boundaries, as $\mathcal{S}_p \left(r\right)=\mathcal{S}^{(o)}\left(r\right)+\mathcal{S}_p^{(sc)}\left(r\right)$. Since $\mathcal{S}^{(o)}\left(r\right)$ is the stress tensor contribution from the homogeneous medium, there is no distinction between the $p=e$ or $p=m$ contributions. 
The homogeneous part of dyadic Green's function ($\overline{\overline{G}}_{e}^{(o)}(\mathbf{r},\mathbf{\tilde{r}})$) on either side of the interface between objects 1 and 2 (as shown in Fig. \ref{fig:schematic}) cancels out, while the scattered part of dyadic Green's function is important, which leads to the dispersion force due to the scattering of fluctuations of electromagnetic fields.

The expressions for $\overline{\overline{G}}_{e}^{(sc)}(\mathbf{r},\mathbf{\tilde{r}})$ when $\mathbf{r}, \mathbf{\tilde{r}} \in V_{2}$ (Fig. \ref{fig:schematic}a) and $\mathbf{r}, \mathbf{\tilde{r}} \in V_{1}$ (Fig. \ref{fig:schematic}b), respectively, are given by
\begin{subequations}
\begin{equation}
\begin{split}
\label{eqn:DGFsc1}
\overline{\overline{G}}_{e}&^{(sc)}(\bm{r},\bm{\tilde{r}}) = ik_2 \sum\limits_{m=-l,\atop l=0}^{l=\infty, \atop m=l} (-1)^{m} \\ & \times \left[ \begin{array}{l}R_{22,l}^{(M)}\bm{M}_{l,m}^{(3)}\left(k_2 \bm{r}\right)\bm{M}_{l,-m}^{(3)}\left(k_2 \bm{\tilde{r}}\right) + \\ R_{22,l}^{(N)} \bm{N}_{l,m}^{(3)}\left(k_2 \bm{r}\right)\bm{N}_{l,-m}^{(3)}\left(k_2 \bm{\tilde{r}}\right) \end{array} \right]
\end{split}
\end{equation}
\begin{equation}
\begin{split}
\label{eqn:DGFsc2}
\overline{\overline{G}}_{e}&^{(sc)}(\bm{r},\bm{\tilde{r}}) = ik_1 \sum\limits_{m=-l,\atop l=0}^{l=\infty, \atop m=l} (-1)^{m} \\ & \times
\left[ \begin{array}{l} R_{11,l}^{(M)}\bm{M}_{l,m}^{(1)}\left(k_1 \bm{r}\right)\bm{M}_{l,-m}^{(1)}\left(k_1 \bm{\tilde{r}}\right) + \\ R_{11,l}^{(N)} \bm{N}_{l,m}^{(1)}\left(k_1 \bm{r}\right)\bm{N}_{l,-m}^{(1)}\left(k_1 \bm{\tilde{r}}\right) \end{array} \right]
\end{split}
\end{equation}
\end{subequations}
where, $R_{ij,l}^{(U)}$ is the Mie refection coefficient due to the source in region $i$ to the observation in region $j$ due to $U=\bm{M,N}$ waves. When $|\bm{r}| > |\bm{\tilde{r}}|$. $k_i=\omega \sqrt{\varepsilon_i(\omega)\mu_i(\omega)}/c$ is the wavenumber in region $i$ ($i=1,2,3$), $\bm{M}_{l,m}^{(p)}(k_i r)$ and $\bm{N}_{l,m}^{(p)}(k_i r)$ are vector spherical wave functions of order $(l,m)$, and superscript $(p)$ refers to the radial behavior of the waves, given by \cite{narayanaswamy2008thermal}
\begin{subequations}
\begin{equation}
\label{eqn:Mvsw}
\bm{M}_{l,m}^{(p)}(k_i r)= \frac{z_l^{(p)}(k_i r)}{\sqrt{l(l+1)}}\left( \hat{\theta}\frac{im Y_{l,m}}{\sin \theta} -\hat{\phi}\frac{\partial Y_{l,m}}{\partial \theta} \right) 
\end{equation}

\begin{equation}
\label{eqn:Nvsw}
\begin{split}
\bm{N}_{l,m}^{(p)}(k_i r)=&\hat{r}\frac{z_l^{(p)}(k_i r)}{k_i r}\sqrt{l(l+1)}Y_{l,m} \nonumber \\ &+ \frac{\xi_l^{(p)}(k_i r)}{\sqrt{l(l+1)}}\left(\hat{\theta}\frac{\partial Y_{l,m}}{\partial \theta} + \hat{\phi}\frac{im Y_{l,m}}{\sin \theta}\right) 
\end{split}
\end{equation}
\end{subequations}

For superscript $p=1$, both $\bm{M}$ and $\bm{N}$ waves are regular vector spherical waves and $z_l^{(1)}$ is the spherical Bessel function of first kind of order $l$. For superscript $p=3$, both $\bm{M}$ and $\bm{N}$ waves are outgoing spherical waves and $z_l^{(3)}$ is the spherical Hankel function of first kind of order $l$. $\xi_l^{(p)}$ is first derivative of Bessel (Hankel) function, defined as $x\xi_l^{(p)}(x)=\displaystyle\frac{d}{dx}\left[x z_l^{(p)}(x)\right]$. $Y_{l,m}$ is the spherical harmonic of order $(l,m)$.

\section{\label{sec:discussion}Results and Discussion}

The spherical configurations of interest in this paper are shown in Fig. \ref{fig:schematic}. To determine the generalized transmissivity for radiative energy transport (Eq. \ref{eqn:generalizedtransmissivity1}) and van der Waals stress (Eq. \ref{eqn:stensor}) for these geometries, we need to find the corresponding electric and/or magnetic dyadic Green's functions, which are expressed in terms of the Mie reflection coefficients. Here, we list boundary conditions of electromagnetic fields to be used to calculate the Mie reflection and transmission coefficients for the four cases: (a) a sphere of radius $r$, (b) a bubble of radius $r$, (c) a spherical shell of interior and exterior radii $r_1$ and $r_2$, and (d) a sphere of radius $r_1$ with a coating of thickness $r_2-r_1$ (the same as case (c) in mathematics).

\begin{widetext}
Case (a): A sphere of radius $r$. Radius $r_+=r+\delta$ and $\delta\rightarrow 0$. Thermal sources outside the sphere, $r \in V_2$. The boundary conditions are given by
\begin{equation}
\label{eqn:sphere}
\begin{cases}
\mu_2 z_l^{(1)}(k_2 r)+R_{22,l}^{(M)}\mu_2 z_l^{(3)}(k_2 r) = T_{21,l}^{(M)} \mu_1 z_l^{(1)}(k_1 r) \\
k_2 \xi_l^{(1)}(k_2 r)+R_{22,l}^{(M)}k_2 \xi_l^{(3)}(k_2 r) = T_{21,l}^{(M)} k_1 \xi_l^{(1)}(k_1 r) 
\end{cases}
\end{equation}

Case (b): A bubble of radius $r$. Radius $r_-=r-\delta$ and $\delta\rightarrow 0$. Thermal sources inside the bubble, $r \in V_1$.
\begin{equation}
\label{eqn:bubble}
\begin{cases}
\mu_1 z_l^{(3)}(k_1 r)+R_{11,l}^{(M)}\mu_1 z_l^{(1)}(k_1 r) = T_{12,l}^{(M)} \mu_2 z_l^{(3)}(k_2 r) \\
k_1 \xi_l^{(3)}(k_1 r)+R_{11,l}^{(M)}k_1 \xi_l^{(1)}(k_1 r) = T_{12,l}^{(M)} k_2 \xi_l^{(3)}(k_2 r) 
\end{cases}
\end{equation}

Case (c): A spherical shell of interior radius $r_1$ and exterior radius $r_2$.
(or Case (d): A sphere of radius $r_1$ coated with a thin film of thickness $r_2-r_1$.

Consider thermal sources in material 1, $r\in V_1$. At $r=r_1$,
\begin{equation}
\label{eqn:shellV1r1}
\begin{cases}
\mu_1 z_l^{(3)}(k_1 r_1)+{R}_{11,l}^{(M)}\mu_1 z_l^{(1)}(k_1 r_1) = \widetilde{R}_{12,l}^{(M)} \mu_2 z_l^{(3)}(k_2 r_1)+\widetilde{\widetilde{R}}_{12,l}^{(M)} \mu_2 z_l^{(1)}(k_2 r_1) \\
k_1 \xi_l^{(3)}(k_1 r_1)+{R}_{11,l}^{(M)}k_1 \xi_l^{(1)}(k_1 r_1) = \widetilde{R}_{12,l}^{(M)} k_2 \xi_l^{(3)}(k_2 r_1)+\widetilde{\widetilde{R}}_{12,l}^{(M)} k_2 \xi_l^{(1)}(k_2 r_1) 
\end{cases}
\end{equation}
and at $r=r_2$,
\begin{equation}
\label{eqn:shellV1r2}
\begin{cases}
\widetilde{R}_{12,l}^{(M)} \mu_2 z_l^{(3)}(k_2 r_2)+\widetilde{\widetilde{R}}_{12,l}^{(M)} \mu_2 z_l^{(1)}(k_2 r_2) = T_{23,l}^{(M)} \mu_3 z_l^{(3)}(k_3 r_2)\\
\widetilde{R}_{12,l}^{(M)} k_2 \xi_l^{(3)}(k_2 r_2)+\widetilde{\widetilde{R}}_{12,l}^{(M)} k_2 \xi_l^{(1)}(k_2 r_2) = T_{23,l}^{(M)} k_3 \xi_l^{(3)}(k_3 r_2)
\end{cases}
\end{equation}

And, consider thermal sources in material 3, $r\in V_3$. At $r=r_2$,
\begin{equation}
\label{eqn:shellV3r2}
\begin{cases}
\mu_3 z_l^{(1)}(k_3 r_2)+{R}_{33,l}^{(M)}\mu_3 z_l^{(3)}(k_3 r_2) = \widetilde{R}_{32,l}^{(M)} \mu_2 z_l^{(1)}(k_2 r_2)+\widetilde{\widetilde{R}}_{32,l}^{(M)} \mu_2 z_l^{(3)}(k_2 r_2) \\
k_3 \xi_l^{(1)}(k_3 r_2)+{R}_{33,l}^{(M)}k_3 \xi_l^{(3)}(k_3 r_2) = \widetilde{R}_{32,l}^{(M)} k_2 \xi_l^{(1)}(k_2 r_2)+\widetilde{\widetilde{R}}_{32,l}^{(M)} k_2 \xi_l^{(3)}(k_2 r_2) 
\end{cases}
\end{equation}
and at $r=r_1$,
\begin{equation}
\label{eqn:shellV3r1}
\begin{cases}
\widetilde{R}_{32,l}^{(M)} \mu_2 z_l^{(1)}(k_2 r_1)+\widetilde{\widetilde{R}}_{32,l}^{(M)} \mu_2 z_l^{(3)}(k_2 r_1) = T_{21,l}^{(M)} \mu_1 z_l^{(1)}(k_1 r_1)\\
\widetilde{R}_{32,l}^{(M)} k_2 \xi_l^{(1)}(k_2 r_1)+\widetilde{\widetilde{R}}_{32,l}^{(M)} k_2 \xi_l^{(3)}(k_2 r_1) = T_{21,l}^{(M)} k_1 \xi_l^{(1)}(k_1 r_1)
\end{cases}
\end{equation}
\end{widetext}
Here, $R_{ij,l}^{(U)}$ and $T_{ij}^{(U)}$ are the Mie reflection and transmission coefficients due to $U$($=\bm{M},\bm{N}$) waves from region $i$ to region $j$, respectively. 

For example, the Mie coefficients due to $\bm{M}$ waves $R_{11,l}^{(M)} $ and $T_{12,l}^{(M)} $ in Case (a) can be obtained from Eq. \ref{eqn:sphere} as
\begin{subequations}
\begin{equation}
\label{eqn:R22}
R_{22,l}^{(M)} = -\frac{\displaystyle\left(\frac{k_2}{\mu_2}\frac{\xi_l^{(1)}(k_2 a)}{z_l^{(1)}(k_2 r)}-\frac{k_1}{\mu_1}\frac{\xi_l^{(1)}(k_1 r)}{z_l^{(1)}(k_1 a)}\right)}{\displaystyle \left(\frac{k_2}{\mu_2}\frac{\xi_l^{(3)}(k_2 a)}{z_l^{(3)}(k_2 r)}-\frac{k_1}{\mu_1}\frac{\xi_l^{(1)}(k_1 r)}{z_l^{(1)}(k_1 a)}\right)}\frac{z_l^{(1)}(k_2 r)}{z_l^{(3)}(k_2 r)}
\end{equation}
\begin{equation}
\label{eqn:T21}
T_{21,l}^{(M)} = \frac{\mu_2}{\mu_1}\left(\frac{y_l^{(1)}(k_2r)}{y_l^{(1)}(k_1 r)}+R_{22,l}^{(M)}\frac{y_{l}^{(3)}(k_2r)}{y_l^{(1)}(k_1 r)}\right)
\end{equation}
\end{subequations}
and the Mie coefficients due to $\bm{M}$ waves $R_{22,l}^{(M)} $ and $T_{21,l}^{(M)} $ in Case (b) can be obtained from Eq. \ref{eqn:bubble} as
\begin{subequations}
\begin{equation}
\label{eqn:R11}
R_{11,l}^{(M)} = -\frac{\displaystyle \left(\frac{k_1}{\mu_1}\frac{\xi_l^{(3)}(k_1 r)}{z_l^{(3)}(k_1 r)}-\frac{k_2}{\mu_2}\frac{\xi_l^{(3)}(k_2 r)}{z_l^{(3)}(k_2 r)}\right)}{\displaystyle\left(\frac{k_1}{\mu_1}\frac{\xi_l^{(1)}(k_1 r)}{z_l^{(1)}(k_1 r)}-\frac{k_2}{\mu_2}\frac{\xi_l^{(3)}(k_2 r)}{z_l^{(3)}(k_2 r)}\right)}\frac{z_l^{(3)}(k_1 r)}{z_l^{(1)}(k_1 r)}
\end{equation}
\begin{equation}
\label{eqn:T12}
T_{12,l}^{(M)} = \frac{\mu_1}{\mu_2}\left(\frac{y_l^{(3)}(k_1r)}{y_l^{(3)}(k_2 r)}+R_{11,l}^{(M)}\frac{y_{l}^{(1)}(k_1r)}{y_l^{(3)}(k_2 r)}\right)
\end{equation}
\end{subequations}
The Mie reflection and transmission coefficients due to $\bm{M}$ waves in Cases (c) and (d) can be obtained by solving Eqs. \ref{eqn:shellV1r1} - \ref{eqn:shellV3r1}. Similarly, the Mie coefficients due to $\bm{N}$ waves can be obtained by simply switching functions of $z_l$ and $\xi_l$ \cite{narayanaswamy2008thermal}.

\subsection{\label{subsec:emissivity} Emissivity of a spherical body}

Now, the Mie scattering coefficients and dyadic Green's functions of a sphere are known. Substituting Eq. \ref{eqn:R11}, Eq. \ref{eqn:T12}, Eq. \ref{eqn:DGFsc1} and Eq. \ref{eqn:DGFsc2} into Eq. \ref{eqn:generalizedtransmissivity1} yields the transmissivity between a single sphere of radius $r$ and a concentric sphere of infinite large radius $r_2\rightarrow \infty$, which is given by (thermal sources within the sphere $r\in V_1$)
\begin{equation}
\label{eqn:transm}
\begin{split}
\mathcal{T}_{rr}&(\omega) =4\pi r^2\frac{\omega^2}{c^2}\Re\sum\limits_{l=0}^{l=\infty} (-1)^{m}\left(2l+1\right) \\ &\times \left(\vert T_{12,l}^{(M)}\vert^2+\vert T_{12,l}^{(N)}\vert^2 \right) z_l^{(1)}(k_ r) \xi_l^{(1)}(k_1 r)
\end{split}
\end{equation}
Equation \ref{eqn:transm} can be modified for the calculation of the spectral emissivity of a sphere, $\epsilon(\omega)=\left(\hbar\omega/2\pi\right)\coth\left(\hbar\omega/2k_B T\right)\mathcal{T}_{rr}/I_{\omega}$, and $I_{\omega}$, that is Planck's specific intensity of a monochromatic plane of frequency $\omega$, is $I_{\omega} = \hbar\omega^3/\left[4\pi^2 c^2 \left(\exp(\hbar\omega/k_B T)-1\right)\right]$.

\begin{figure}[h]
\centering
\includegraphics[width=6.5 cm]{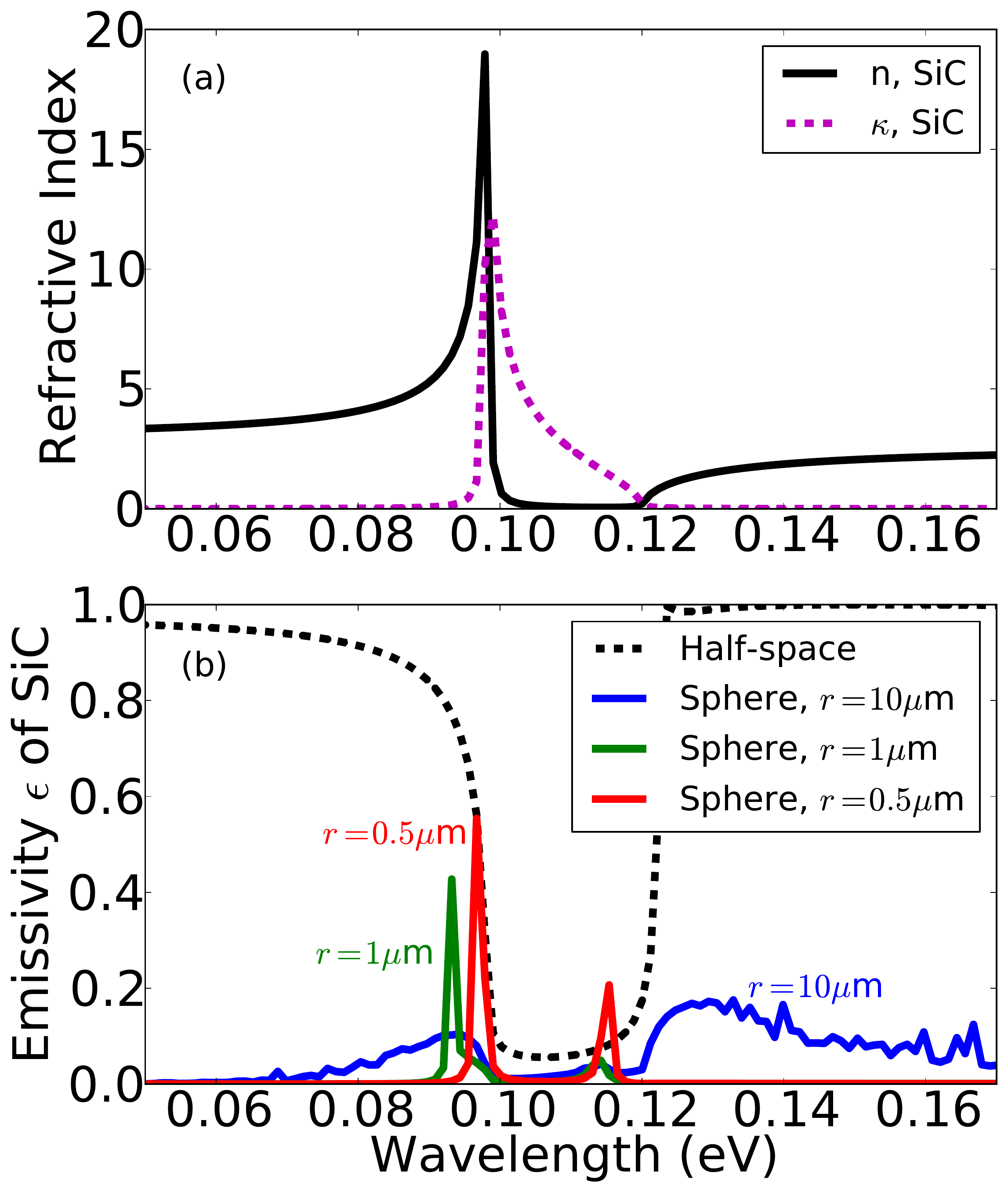}
\caption{\label{fig:emissivity} Refractive index and emissivity of SiC spheres of different sizes. (a) real and imaginar parts of the refractive index, $n+i\kappa$, of SiC, and (b) emissivity of a SiC half-space and SiC spheres of radii 10 $\mu$m, 1 $\mu$m and 0.5 $\mu$m at room temperature $T=300$ K.
}
\end{figure}

Figure \ref{fig:emissivity}a shows the refractive index, $n+i\kappa$, of silicon carbide (SiC). The emissivity is plotted for SiC half-space and spheres of different radii of 10 $\mu$m, 1 $\mu$m and 0.5 $\mu$m at room temperature $T=300$ K in Fig. \ref{fig:emissivity}b. The black dotted line shows the emissivity of a half-space, which has a low emissivity ($\epsilon\approx 0.1$) when refractive index $\kappa \gg n\approx 0$. It can be observed that, as the radius of sphere reduces, the sharp peaks occurs in the emission spectrum. For example, there are two emission peaks at $\omega = 0.098$ eV and 0.114 eV, which are corresponding the frequencies at which the dielectric function reaches its minimum. Another way of explaining these peaks is that the radius of the spheres is much larger than the absorption depth, $d_{abs}=c/2\omega \kappa$, at those wavelengths because of multiple reflections within the object.

\subsection{\label{subsec:vdW}van der Waals surface energy of a spherical bubble}

Substituting the scattered part of spherical dyadic Green's functions Eq. \ref{eqn:DGFsc1} and Eq. \ref{eqn:DGFsc2} into Eq. \ref{eqn:stensor}, it yields the $\hat{r}\hat{r}$ component of the stress tensor at the interior vacuum-medium interface within a bubbe, i.e., at $r \approx r \pm \delta$ as $\delta \rightarrow 0$, which can be written as
\begin{equation}
\label{eqn:stresstensor}
\begin{split}
&\mathcal{S}_{rr}(r) =\Re\int\limits_0^{\infty} d\omega \frac{\hbar\omega^3}{4 \pi^2 c^3}\coth\left(\frac{\hbar \omega}{2k_B T}\right) \sum\limits_{l=0}^{\infty}\left(2l+1\right) \\ &\times \left(R_{11,l}^{(M)}+R_{11,l}^{(N)}\right) \left[\left(1-\frac{l(l+1)}{k_1^2r^2}\right)z_l^{(1)2}(k_1 r)+\xi_l^{(1)2}(k_1 r)\right]
\end{split}
\end{equation}

Let the hydrostatic pressures in the vacuum and medium out of the bubble be $p_{V}$ and $p_{M}$ respectively. Balance of forces at the interface yields a modified Young-Laplace equation, and the pressure change across the interface of a bubble can be expressed below \cite{zheng2014surface},
\begin{subequations}
\begin{equation}
\label{eqn:balance}
\begin{split}
\Delta p=p_V-p_{M}=\mathcal{S}_{rr}(r)+\frac{2\sigma_\infty(T)}{r}
\end{split}
\end{equation}
\begin{equation}
\label{eqn:surfaceenergy}
\begin{split}
\Rightarrow \sigma(r,T)\approx\sigma_{\infty}(T) +\frac{r}{2}\mathcal{S}_{rr}(r)
\end{split}
\end{equation}
\end{subequations}
where $\sigma_{\infty}$ is temperature dependent surface tension. Manipulation of Eq. \ref{eqn:balance} gives rise to a temperature as well as size dependent surface energy $\sigma(r,T)$ of a bubble of radius $r$ at temperature $T$ in Eq. \ref{eqn:surfaceenergy}.

\begin{figure}[h]
\centering
\includegraphics[width=7.5 cm]{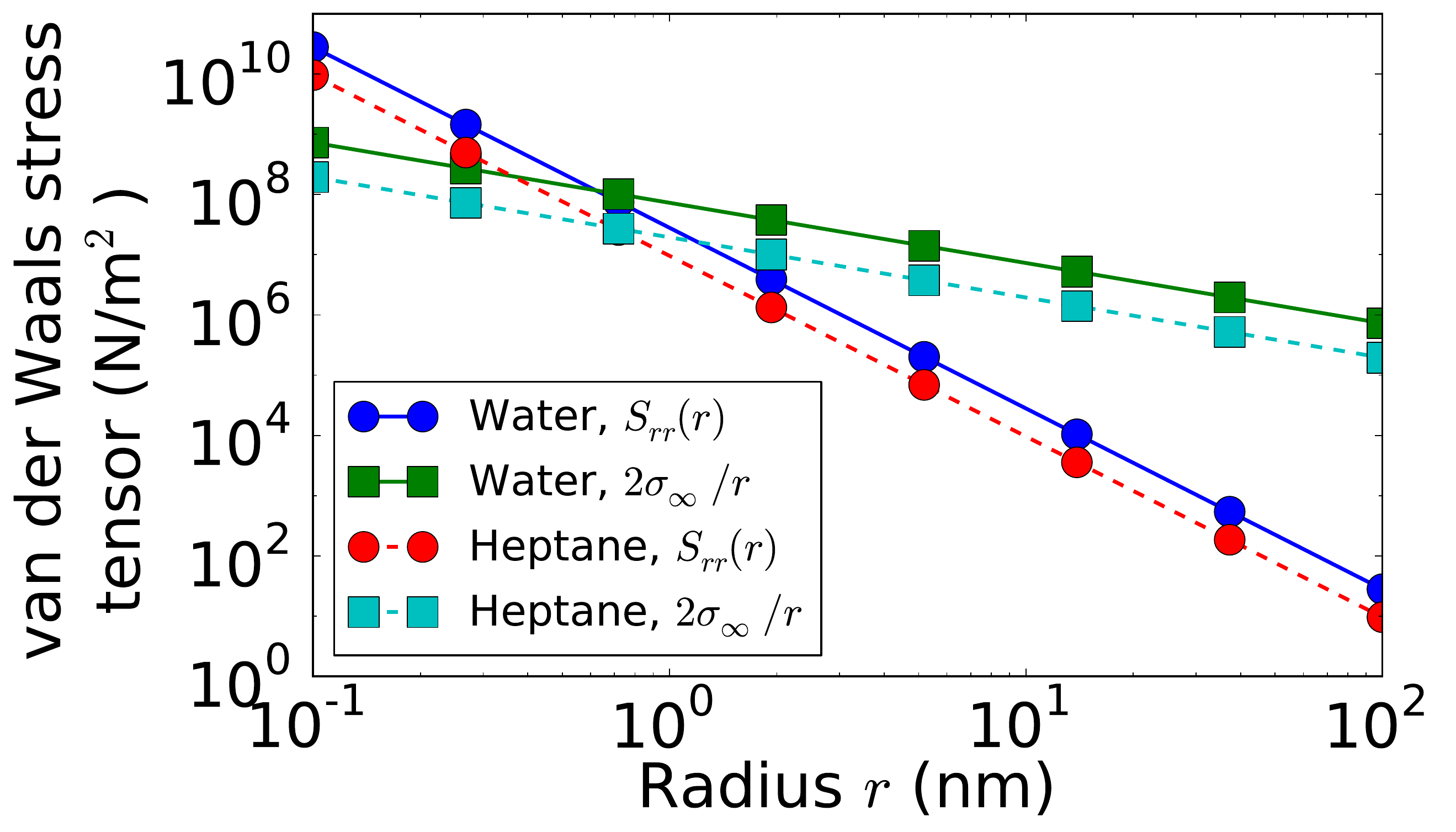}
\caption{\label{fig:surfaceenergy} Size dependence of van der Waals surface stress ($\mathcal{S}_{rr}$) and stress due to surface tension ($\sigma_{\infty}$) of a spherical bubble, in water and heptane, respectively. The intersection of the stresses indicates a critical value of radius $r_{cr}$, which is defined to be such that $\mathcal{S}_{rr}=2\sigma_{\infty}/r$. As $r\gg r_{cr}$, surface tension dominates the surface energy of a bubble in the process of homogeneous nucleation. Whereas $r<r_{cr}$, van der Waals stress plays a significant role in surface energy of a micro and nano-sized bubble.
}
\end{figure}

The stress tensor $\mathcal{S}_{rr}(r)$ and surface stress due to surface tension $2\sigma_{\infty}/r$ are plotted for water and heptane in Fig. \ref{fig:surfaceenergy}. The thermophysical property data can be obtained through the NIST Chemistry WebBook \cite{nistwebbook}. It can be seen that the $\hat{r}\hat{r}$ component of the van der Waals stress obeys a $r^{-3}$ rule, while the stress due to surface tension $2\sigma_{\infty}/r$ follows $r^{-1}$. The interaction of two curves implies a critical radius $r_{cr}$ which is defined to be such that $\mathcal{S}_{rr}=2\sigma_{\infty}/r$. As $r\gg r_{cr}$, surface tension dominates the surface energy of a bubble in the process of homogeneous nucleation, whereas $r<r_{cr}$, van der Waals stress plays a significant role in surface energy of a nano-sized bubble. It can be found that the critical radius is $r_{cr} \approx 0.8$ nm for both water and heptane. Eq. \ref{eqn:surfaceenergy} shows that surface energy $\sigma(r,T)\approx r\mathcal{S}_{rr}(r)/2$ when $r \ll r_{cr}$, and $\sigma(r,T)\approx \sigma_{\infty}(T)$ when $r\gg r_{cr}$ \cite{zheng2014surface}.

\section{\label{sec:sum}Conclusion}

Fluctuations of electromagnetic fields lead to radiative energy and momentum transfer between objects. They can be described by the cross-spectral densities of the electromagnetic fields, which are expressed in terms of the dyadic Green's functions of the vector Helmholtz equation. The radiative energy and momentum transfer between planar objects have been studied during the past decades, while that between other geometries, such as spherical or cylindrical shapes, have not been investigated well. Proximity approximation and/or modified proximity approximation is one commonly used numerical approach to probe the near-field thermal radiation and fluctuation-induced van der Waals pressure between curved surface, but its validity depends strongly on the configurations and sizes of objects.

In this paper, we apply fluctuational electrodynamics directly to evaluate emissivity of thermal radiation and van der Waals contribution to surface energy for various spherical shapes, such as a sphere, a bubble, a spherical shell and a coated sphere, in a homogeneous and isotropic medium. The dyadic Green's function formalism of thermal radiative transfer and van der Waals stress for different spherical configurations have been developed. We have shown the size dependence and wavelength selectivity of emission spectrum and surface energy of the micro/nanoscale spheres for small radii. The emissivity of sharp peaks can be observed at wavelengths corresponding to the characteristics of the material's refractive index, and improved as the size of object decreases. van der Waals contribution dominates the surface energy when size of object is reduced to a nanoscopic length scale. The study of fluctuation-induced radiative energy and momentum transfer for spherical shapes has great applications in the nanoscale engineering, e.g., nanoparticles/beads and coated spherical shells, and it deserves further investigation in future.

\section*{Acknowledgments}
This work is funded by the Start-up Grant through the College of Engineering at the University of Rhode Island.



%

\end{document}